\documentstyle[aps,prl,preprint,floats,epsf]{revtex}

\begin{document}

\preprint{\tighten\vbox{\hbox{\hfil CLNS 97/1480}
                        \hbox{\hfil CLEO 97-9}
}}

\title{Determination of the Michel Parameters and the 
$\tau$ Neutrino \\ Helicity in $\tau$ Decay}

\author{CLEO Collaboration}
\date{\today}

\maketitle
\tighten

\begin{abstract}
Using the CLEO II detector at the $e^+e^-$ storage 
ring CESR, we have determined the Michel parameters 
$\rho$, $\xi$, and $\delta$
in $\tau^\mp\rightarrow l^\mp\nu\overline{\nu}$ decay as well as
the $\tau$ neutrino helicity parameter $h_{\nu_\tau}$ in 
$\tau^\mp\rightarrow\pi^\mp\pi^0\nu$ decay.
From a data sample of $3.02\times 10^6$ $\tau$ pairs produced at
$\sqrt{s}=10.6 \mbox{ GeV}$, using events of the topology
$e^+e^-\rightarrow\tau^+\tau^-\rightarrow%
(l^\pm\nu\overline{\nu})(\pi^\mp\pi^0\nu)$ and 
$e^+e^-\rightarrow\tau^+\tau^-\rightarrow%
(\pi^\pm\pi^0\overline{\nu})(\pi^\mp\pi^0\nu)$, and 
the determined sign of $h_{\nu_\tau}$ 
\cite{albin2,ich}, the combined result of the three
samples is:
$\rho                     =  0.747\pm 0.010\pm 0.006$, 
$\xi                      =  1.007\pm 0.040\pm 0.015$,
$\xi\delta                =  0.745\pm 0.026\pm 0.009$, and 
$h_{\nu_\tau}             = -0.995\pm 0.010\pm 0.003$.
The results are in agreement with the Standard Model 
$V$-$A$ interaction.
\end{abstract}


\newpage

{
\renewcommand{\thefootnote}{\fnsymbol{footnote}}

\begin{center}
J.~P.~Alexander,$^{1}$ C.~Bebek,$^{1}$ B.~E.~Berger,$^{1}$
K.~Berkelman,$^{1}$ K.~Bloom,$^{1}$ D.~G.~Cassel,$^{1}$
H.~A.~Cho,$^{1}$ D.~M.~Coffman,$^{1}$ D.~S.~Crowcroft,$^{1}$
M.~Dickson,$^{1}$
% ...
\end{center}
 
\small
\begin{center}
$^{1}${Cornell University, Ithaca, New York 14853}\\
% ...
$^{13}${University of Oklahoma, Norman, Oklahoma 73019}\\
$^{14}${Purdue University, West Lafayette, Indiana 47907}\\
% ...
\end{center}

\setcounter{footnote}{0}
}
\newpage


\section{INTRODUCTION}

The most general, local, derivative-free, and 
lepton-number-conserving four fermion point interaction 
\cite{michel,bouch,okun} for leptonic
$\tau$ decays yields in the helicity projection form 
\cite{scheck} the following matrix element
\begin{equation}
\label{general}
{\cal M}  = 4 \frac{G_l}{\sqrt{2}}
\sum_{ \stackrel{ \gamma = S,V,T }{ \epsilon,\mu = R,L}}
g^\gamma_{\epsilon\mu}
\left[ 
\overline{u}_\epsilon (l^-) \Gamma_\gamma 
v_j( \overline{\nu}_l ) 
\right]
\left[ 
\overline{u}_i (\nu_\tau ) \Gamma^\gamma
u_\mu(\tau^-) 
\right] \, ,
\end{equation} 
where $G_l$ parametrizes the total strength of the 
interaction. The matrices $\Gamma_\gamma$ define the
properties of the two currents under a Lorentz
transformation with $\gamma = S,V,T$ for
scalar, vector, and tensor interactions. 
The indices $\epsilon$ and $\mu$ label the right- or 
lefthandedness (R,L) of the charged leptons. 
For a given $\epsilon$, $\mu$, and $\gamma$,
the handedness of the neutrinos labeled by $j$ and $i$
are fixed. Only ten of the twelve
complex coupling constants $g^\gamma_{\epsilon\mu}$ 
are linearly independent. 
In the Standard Model $V$-$A$ interaction,
the only non-zero coupling constant is $g^V_{LL}=1$.

The interaction described by Eqn.~\ref{general} is fully 
determined by 19 real parameters. 
Without measuring the neutrinos and
the spin of the outgoing charged lepton, only the four
Michel parameters \cite{michel,bouch,okun}
$\rho$, $\eta$, $\xi$, and $\delta$ are experimentally accessible.
They are bilinear combinations of the coupling constants
$g^\gamma_{\epsilon\mu}$  and
appear in the predicted energy spectrum of the 
charged lepton $l^\mp$ emitted in the decay
$\tau^\mp\rightarrow l^\mp\nu\overline{\nu}$. 
In the $\tau$ rest frame, neglecting radiative corrections
and terms proportional to $m^2_l/m^2_\tau$, this
spectrum is given by
\begin{eqnarray}  
\label{michel}
\frac{ d\Gamma (\tau^\mp\rightarrow l^\mp\nu\overline{\nu})} 
{d\Omega dx} & = &
\frac{G^2_F m^5_\tau}{192\pi^4} x^2 
\bigg[
3(1-x) + \frac{2}{3}\rho (4x-3) + 6\eta
\frac{m_l}{m_\tau}\frac{1-x}{x} \nonumber\\ 
& &
\mp\xi P_\tau\cos\theta\left( (1-x) + \frac{2}{3}\delta
(4x-3)\right)\bigg] \, ,
\end{eqnarray}
where $x=2E_l/m_\tau$ is the scaled charged lepton energy, 
$P_\tau$ the $\tau$ polarization, and $\theta$ the angle 
between $\tau$ spin and lepton momentum. 
In the Standard Model the $V$-$A$ charged weak current
is characterized  by $\rho = 3/4$, $\eta = 0$, 
$\xi = 1$ and $\delta = 3/4$.

A measurement of the Michel parameters allows one to limit
the coupling constants $g^\gamma_{\epsilon\mu}$.
For example $\xi$ and $\delta$ determine the probability,
$P^\tau_R$, for a righthanded $\tau$ lepton to participate
in leptonic $\tau$ decays:
\begin{equation}
\label{rechts}
P^\tau_R= 
\frac{1}{4}{\vert g^S_{RR} \vert}^2 + 
\frac{1}{4}{\vert g^S_{LR} \vert}^2
          +{\vert g^V_{RR} \vert}^2 +  
           {\vert g^V_{LR} \vert}^2
         +3{\vert g^T_{LR} \vert}^2 =
\frac{1}{2}\left[1+\frac{1}{9}\left(3\xi-16\xi\delta\right)\right]
\end{equation}

Despite the progress made in recent years
\cite{albin2,ich,rusky,klaus,sld,holger,alf2,%
ivan,alf3,l3,mandeepa,sld97},
the determination of the space-time 
structure in leptonic and semihadronic $\tau$ decays
is still an order of magnitude less precise than 
in $\mu$ decay, indicating a need for 
high precision measurements of the
Michel parameters in leptonic $\tau$ decays
as well as of the $\tau$ neutrino helicity
in semihadronic $\tau$ decays.
In this publication, we present measurements of
$\rho$, $\xi$, $\delta$, and the $\tau$ neutrino helicity
$h_{\nu_\tau}$ from an analysis of  
$e^+ e^- \rightarrow\tau^+\tau^-\rightarrow
(l^\pm\nu\overline{\nu})(\pi^\mp\pi^0\nu)$ 
and $(\pi^\pm\pi^0\overline{\nu})(\pi^\mp\pi^0\nu)$
events. The parameter 
$h_{\nu_\tau}$ is given by 
$h_{\nu_\tau} = 2g_V g_A /(g^2_V + g^2_A)$, where
$g_V$ and $g_A$ are the vector and axialvector couplings.
In the Standard Model, with purely lefthanded
neutrinos, one expects $h_{\nu_\tau}=-1$.

The $(l^\pm\nu\overline{\nu})(\pi^\mp\pi^0\nu)$ sample
used here is correlated with that of ref.~\cite{mandeepa}.
There, the Michel parameters $\rho$ and $\eta$ have been determined
with emphasis on a precise measurement of $\eta$, in which 
the sensitivity comes mainly from the low momentum part
of the muon spectrum.
Here, the emphasis lies on the determination of the
spin dependent Michel parameters $\xi$ and $\delta$.


\section{METHOD OF THE MEASUREMENT}

Eqn.~\ref{michel} shows that the measurement of $\xi$
and $\delta$ requires the knowledge of the $\tau$ spin
orientation. In $e^+e^-$ annihilation at 
$\sqrt{s}\approx 10\mbox{ GeV}$
the average $\tau$ polarization is zero and no information
on $\xi$ and $\delta$ can be obtained from
single $\tau$ decays. However, 
spin-spin correlations exist between the two $\tau$ leptons 
in $e^+e^-\rightarrow\tau^+\tau^-$, leading to 
correlations between kinematical properties of
the decay products.
These correlations have been used before 
\cite{ich,klaus,holger} for the determination of $\xi$, $\delta$,
and $h_{\nu_\tau}$, where leptonic as well as semihadronic
decays served as spin analyzers. Here we use the 
semihadronic $\tau$ decay $\tau^\mp\rightarrow\pi^\mp\pi^0\nu$ 
as spin analyzer. Its advantages are a large branching ratio,
a very well understood hadronic current,
and an experimentally clean signal.

In the Born approximation, the matrix element 
for the differential cross section of  
$e^+e^-\rightarrow\tau^+\tau^-\rightarrow%
(l^\pm\nu\overline{\nu})(\pi^\mp\pi^0\nu)$ has, 
after integration over the unobserved neutrino
degrees of freedom and summation over unobserved spins,
the following structure (see for example ref.~\cite{was1}):
\begin{equation}
\label{gesamt}
{\vert {\cal M} \vert}^2  = 
H   P    [L_1 + \rho L_2 + \eta   L_3  ] + 
h_{\nu_\tau} H^\prime_\alpha   C^{\alpha\beta}          
[\xi L^\prime_{1\beta} + \xi\delta L^\prime_{2\beta} ] \, .
\end{equation}

The first term is the spin averaged part of the
differential cross section. The second
term contains the spin correlation.
For the semihadronic decay 
$\tau^\mp\rightarrow\pi^\mp\pi^0\nu$
the matrix element is formulated 
for an arbitrary mixing of $V$ and $A$ couplings,
parametrized by the neutrino helicity $h_{\nu_\tau}$. 
The spin averaged part of this matrix element
is indicated by $H$ and the spin dependent part by $H^\prime$. 
The symbols $L_i$ and $L_i^\prime$ 
are the Lorentz invariant formulations
of the corresponding terms in Eqn.~\ref{michel}. 
The spin averaged $\tau$ pair production 
is denoted by $P$ and the production spin correlation 
matrix by $C^{\alpha\beta}$. 

The spin analyzer $\tau^\mp\rightarrow\pi^\mp\pi^0\nu$
can resolve the ratio of longitudinal to transverse
polarization of the intermediate $\rho$ meson, but,
because of the absence of interference terms, 
cannot separate its transverse polarization into the left
and righthanded part. 
Thus, our  
$(l^\pm\nu\overline{\nu})(\pi^\mp\pi^0\nu)$ events
are sensitive to $\rho$, $\eta$, $h_{\nu_\tau}\xi$,
and $h_{\nu_\tau}\xi\delta$ (see Eqn.~\ref{gesamt}), whereas 
our $(\pi^\pm\pi^0\overline{\nu})(\pi^\mp\pi^0\nu)$
events allow a determination of the product $h_{\nu_\tau}^2$.
The signs of $\xi$ and $h_{\nu_\tau}$ are well known
from other experiments \cite{albin2,ich} and no attempt
is made in this analysis to remeasure these signs.

Not all of the kinematical quantities 
needed to evaluate Eqn.~\ref{gesamt} for each detected event
are well determined.
For example, the azimuthal
angle of the $\tau$ momentum around the two-pion momentum 
can take on a range of values, restricted by 
kinematical constraints. Initial state radiation, radiative 
corrections to the decays $\tau^\mp\rightarrow l^\mp \overline{\nu}\nu$
and $\tau^\mp\rightarrow\pi^\mp\pi^0\nu$,
external bremsstrahlung, and uncertainties of the
measured momenta will also modify the evaluation of
Eqn.~\ref{gesamt}.

These indeterminancies are taken into account 
in a likelihood function by forming a weighted
sum over all possible kinematical configurations.
The weights are derived 
by assuming that the radiative effects and the resolution of the detector
factorize from the Born level matrix element and do not depend
on the fit parameters.
Formally, the likelihood function is taken to be
\begin{equation}
\label{likel}
L(\Theta\vert\vec{a}) := P(\vec{a}\vert\Theta ) =
\frac{%
\int 
{\vert {\cal M}(\vec{\alpha} ,\vec{\beta}\vert\Theta ) \vert}^2 
\eta(\vec{\alpha}) w(\vec{a},\vec{\beta} \vert \vec{\alpha})
d\vec{\beta}d\vec{\alpha}  
}%
{%
\int 
{\vert {\cal M}
(\vec{\alpha},\vec{\beta}\vert\Theta )  
\vert}^2 \eta(\vec{\alpha})
w(\vec{a},\vec{\beta} \vert \vec{\alpha})
d\vec{\beta} d\vec{\alpha} d\vec{a}
} \, , 
\end{equation}
where ${\vert {\cal M}\vert}^2$  
is given by Eqn.~\ref{gesamt}. 
The vector $\Theta$ represents the set of parameters 
$(\rho , \eta , h_{\nu_\tau}\xi , h_{\nu_\tau}\xi\delta )$ 
that are determined in the fit.
As discussed below, we also include the $\Theta$ dependence
of all significant sources of background in the event likelihood.
The vector $\vec{a}$ contains all measured quantities, 
{\it i.e.}~the momenta of the charged lepton and the two pions.
The vector $\vec{\beta}$ contains all unmeasured quantities, 
such as those associated with the neutrinos, the photons 
of the initial state bremsstrahlung which
mostly escape undetected down the beam pipe, radiated
photons in the decay, and photons from external bremsstrahlung.
The vector $\vec{\alpha}$ represents the value of
the measured quantities before resolution and radiative effects.
The weight $w(\vec{a},\vec{\beta} \vert \vec{\alpha})$ contains 
all of the resolution and radiative effects, and
the integrals over $\vec{\alpha}$ perform convolutions with
the detector resolution.
The acceptance function of the detector is denoted by $\eta$ 
and depends only on $\vec{\alpha}$.
The denominator in Eqn.~\ref{likel} ensures that
the likelihood 
integrated over $\vec{a}$ is normalized to unity for all 
values of $\Theta$.

The integration over the unmeasured quantities is done 
analytically as far as possible. The remaining integration is
performed numerically following a hit or miss approach for
the relevant kinematical variables.
Hence, a certain number of trials is done for each observed event
to generate the unmeasured quantities $\vec{\beta}$
(radiated photons),
and the ``before radiation and detector resolution'' 
values $\vec{\alpha}$ of the measured quantities,
under the hypothesis that the event is a $\tau$ pair.
The fraction of trials which are successful, $f_{hit}$, 
is a measure of the goodness of the hypothesis.
The Monte Carlo integration over the
measured quantities in  the denominator of
Eqn.~\ref{likel} is done with a full detector simulation.
This technique was used in ref.~\cite{ich}. Here
we have applied only minor changes, such as, for example, 
taking the radiative corrections in the semihadronic 
decay into account.
A full description can be found in ref.~\cite{dokto}.

The effectiveness of this technique has been demonstrated
by generating events with the KORALB/TAUOLA 
\cite{koral,tauola} Monte Carlo program and applying 
the fit method to these events. 
The results are compatible with the input within the
statistical errors of the test, which are of order
$0.01\%$. These tests have been performed
for Standard Model input values as well as for non-Standard 
Model values.
Thus, at the level of accuracy needed here, we have
demonstrated that the method is unbiased, and 
the factorization assumption
mentioned above is justified.


\section{DATA SELECTION}

The measurements presented here were performed with the
CLEO II detector \cite{cleodek} at the $e^+e^-$ storage 
ring CESR.
The data sample used here was collected in the years
between $1990$ and $1994$ at center of mass energies
around $\sqrt{s}=10 \mbox{ GeV}$.
The integrated luminosity is $\approx 3.5 \mbox{ fb}^{-1}$,
with about $3.02\times 10^6$ $\tau$ pairs produced.

Events with exactly two charged tracks 
with a charge sum of zero are selected. 
Each track must have a momentum 
greater than $500 \mbox{ MeV/c}$ and its distance 
of closest approach to the interaction point in 
the plane transverse to the beam must be less 
than $10\mbox{ mm}$.
The polar angle of each track relative to the beam  
must fulfill 
the condition $\vert\cos\theta\vert < 0.71$.
The two tracks are required to be
separated by an opening angle of more than $90^\circ$.

To suppress non $\tau$ background we require that not more than
one of the two tracks has a momentum greater than
$85\%$ of the beam energy. 
The total visible energy in the electromagnetic calorimeter 
has to be greater than $20\%$  and less than $85\%$ of 
$\sqrt{s}$. Additionally, the momenta $\vec{p}_i$ of
the two tracks have to fulfill the condition
$\vert\vec{p}_1 + \vec{p}_2 \vert / (\vert\vec{p}_1\vert +
\vert\vec{p}_2\vert)>0.05$ to suppress cosmic rays.

Photons used to reconstruct $\pi^0$ mesons are defined as 
calorimeter showers that are not matched to a charged track,
with an energy greater than $50\mbox{ MeV}$ and
a polar angle of $\vert\cos\theta\vert < 0.71$.
To veto against feed across from other
$\tau$ decays, we require zero photon-like isolated showers 
(more than $30\mbox{ cm}$ from the closest track
projection into the calorimeter)
with an energy of more than $75\mbox{ MeV}$ for polar angles of
$\vert\cos\theta\vert < 0.71$ and an energy of more
than $100\mbox{ MeV}$ in the case 
of $\vert\cos\theta\vert > 0.71$.

In the lepton-versus-$\rho$ sample exactly one $\pi^0$ is
required with 
$-4 < (m_{\gamma\gamma}-m_{\pi^0})/\sigma_{m_{\gamma\gamma}} < 3$,
where the mass resolution $\sigma_{m_{\gamma\gamma}}$ 
is typically between 
$5\mbox{ MeV/c}^2$ and $10\mbox{ MeV/c}^2$ depending
on the $\pi^0$ energy.
The momentum of the reconstructed $\pi^0$ has to
be greater than $300\mbox{ MeV/c}$.

The track further away in angle from the reconstructed 
$\pi^0$ is required to be either an electron or a 
muon. Tracks are identified as electrons when their momentum 
and $dE/dx$ information from the tracking system, as well 
as the energy measurement in the electromagnetic calorimeter, 
are consistent with the electron hypothesis.
Tracks with momenta greater than $1.5\mbox{ GeV/c}$ 
are identified as muons if they 
match to hits in the muon counters
beyond at least three absorption lengths of material.

The invariant mass of the reconstructed $\pi^0$
and the charged pion candidate has to satisfy 
$m_{\pi^\mp\pi^0}> 0.5\mbox{ GeV/c}^2$.
The missing mass $m_{miss}$ of the event
has to fulfill the condition $m_{miss} > 0.1 \sqrt{s}$.
Additionally, we require the Monte Carlo success rate
$f_{hit}$ (see previous section) to be at least $>6.6\%$.
After these three cuts the remaining background from 
Bhabha events, two-photon interactions, 
and $q\overline{q}$ events is negligible.

In the $\rho$-versus-$\rho$ sample exactly 
two $\pi^0$ mesons are required with
$-4 < (m_{\gamma\gamma}-m_{\pi^0})/\sigma_{m_{\gamma\gamma}} < 3$.
A momentum cut of greater than $300\mbox{ MeV/c}$ is applied 
on the reconstructed $\pi^0$ mesons.
The $\pi^0$ mesons are associated with the charged tracks
by their nearness in angle. 
The invariant mass $m_{\pi^\mp\pi^0}$
of the $\rho$ meson candidates have to be greater than
$0.5\mbox{ GeV/c}^2$. These cuts are identical to the
$l$-vs-$\rho$ selection and suppress
feed across from other $\tau$ events. In addition,
the missing transverse momentum
of the event, $p_T$, and $E_{TOT}$, the total visible 
energy of the observed
particles, have to satisfy $p_T /(\sqrt{s} - E_{TOT})>0.1$,
$p_T/(\sqrt{s}/2) >0.075$, and $E_{TOT}/\sqrt{s}>0.3$.
We again require $f_{hit} > 6.6\%$. 
After these cuts the contribution from non $\tau$
background is negligible.

\begin{table}
\caption{Background contribution from other $\tau$ decays.}
\label{back}
\begin{center}
\begin{minipage}[t]{5.4truecm}
\begin{center}
\begin{tabular}{l l}
\multicolumn{2}{c}{electrons 33531 accepted events}   \\ \hline
                                         & estimated     \\
\raisebox{2.0ex}[-1.5ex]{event topology} & backgr.~$\%$  \\ \hline
$(e^\pm\nu\overline{\nu})  
(\pi^\mp\pi^0\pi^0\nu)            $ & $ 1.78  \pm 0.20 $ \\ 
$(e^\pm\nu\overline{\nu})
(K^\mp\pi^0\nu)                   $ & $ 1.94  \pm 0.20 $ \\ 
$(\pi^\pm\overline{\nu}) 
(\pi^\mp \pi^0\nu)                $ & $ 0.14  \pm 0.03 $ \\ 
$(e^\pm\nu\overline{\nu})
(\pi^\mp\nu)                      $ & $ 0.13  \pm 0.02 $ \\ 
remaining sources                   & $ 0.96  \pm 0.10 $ \\ 
\hline
$ \Sigma                          $ & $ 4.95 \pm 0.30 $ \\
\end{tabular} 
\end{center}
\end{minipage}
\hfill
\begin{minipage}[t]{5.4truecm}
\begin{center}
\begin{tabular}{l l}
\multicolumn{2}{c}{muons, 21680 accepted events}       \\ \hline
                                         & estimated     \\
\raisebox{2.0ex}[-1.5ex]{event topology} & backgr.~$\%$  \\ \hline
$(\mu^\pm\nu\overline{\nu}) 
(\pi^\mp\pi^0\pi^0\nu)            $ & $ 1.73   \pm 0.20 $  \\
$(\mu^\pm\nu\overline{\nu}) 
(K^\mp\pi^0\nu)                   $ & $ 2.00   \pm 0.20 $  \\
$(\pi^\pm\overline{\nu}) 
(\pi^\mp \pi^0  \nu)              $ & $ 1.29   \pm 0.18 $  \\
$(\mu^\pm\nu\overline{\nu}) 
(\pi^\mp\nu)                      $ & $ 0.14   \pm 0.03 $  \\
remaining sources                   & $ 0.90   \pm 0.10 $  \\ 
\hline
$ \Sigma                          $ & $ 6.06  \pm 0.35 $  \\ 
\end{tabular} 
\end{center} 
\end{minipage}
\hfill
\begin{minipage}[t]{5.4truecm}
\begin{center}
\begin{tabular}{l l}
\multicolumn{2}{c}{$\rho$ mesons, 11177 accepted events} \\ \hline
                                         & estimated     \\
\raisebox{2.0ex}[-1.5ex]{event topology} & backgr.~$\%$  \\ \hline
$(\pi^\pm\pi^0\overline{\nu}) 
(\pi^\mp\pi^0\pi^0\nu)            $ & $ 3.95  \pm 0.45 $   \\
$(\pi^\pm\pi^0\overline{\nu}) 
(K^\mp\pi^0\nu)                   $ & $ 4.31  \pm 0.50 $   \\
                                    &                      \\
                                    &                      \\
remaining sources                   & $ 2.04  \pm 0.20 $   \\ 
\hline
$ \Sigma                          $ & $ 10.30 \pm 0.70 $    \\ 
\end{tabular} 
\end{center} 
\end{minipage}
\end{center}
\end{table}

This selection results in a sample of $66388$ accepted events,
comprising $33531$ candidates in the topology
$(e^\mp \nu\overline{\nu})(\pi^\pm\pi^0 \overline{\nu})$,
$21680$ candidates in
$(\mu^\mp \nu\overline{\nu})(\pi^\pm\pi^0 \overline{\nu})$, and
$11177$ candidates in 
$(\pi^\mp\pi^0 \nu)(\pi^\pm\pi^0 \overline{\nu})$.
These numbers of events are in good agreement 
with expectations based on world
average branching ratios \cite{pdg}.

In all three samples, $e$-vs-$\rho$, $\mu$-vs-$\rho$, 
and $\rho$-vs-$\rho$, the
background from non $\tau$ events is insignificant.
The background from $\tau$ events is estimated
by using the KORALB/TAUOLA Monte Carlo \cite{koral,tauola}.
The results are listed in Table \ref{back}, where the
errors reflect statistical, experimental, and theoretical
uncertainties. The total contribution from $\tau$ background 
is about $5\%$ for 
$(e^\mp \nu\overline{\nu})(\pi^\pm\pi^0 \overline{\nu})$,
around $6\%$ for 
$(\mu^\mp \nu\overline{\nu})(\pi^\pm\pi^0 \overline{\nu})$, 
and approximately $10\%$ for
$(\pi^\mp\pi^0 \nu)(\pi^\pm\pi^0 \overline{\nu})$ events.
The ``remaining sources'' (Table \ref{back}) are from a
variety of modes.


\section{DATA ANALYSIS}

\begin{figure}[thb]
\begin{center}
\leavevmode
\unitlength1.0cm
\epsfxsize=16.5cm
\epsfysize=6.0cm
\epsffile{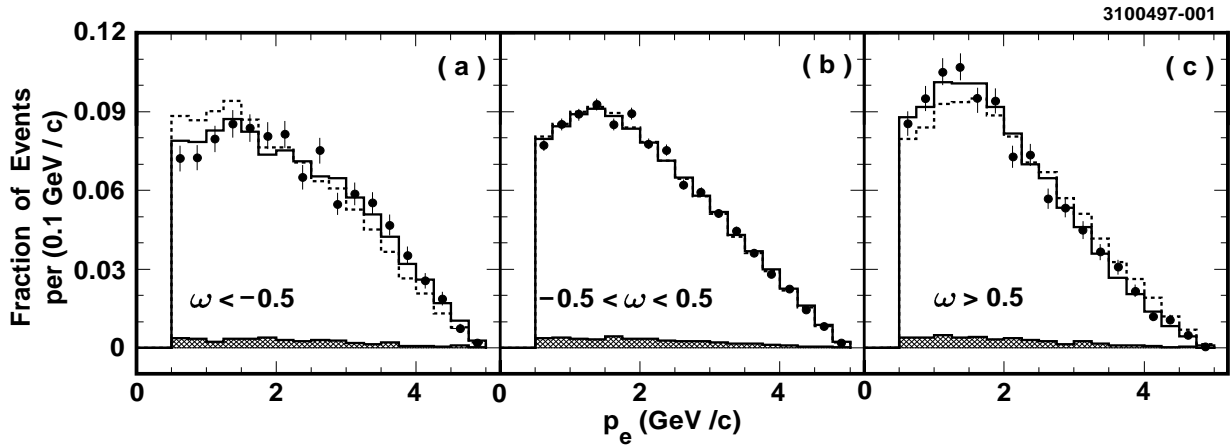}
\caption{\label{sens_elk}
Electron momentum spectrum in different regions of $\omega$.
The variable $\omega$ is sensitive to the spin of the
$\tau$ lepton in the decay 
$\tau^\mp\rightarrow\pi^\mp\pi^0\nu$ 
($\omega < 0 \Rightarrow
\tau^-$ lefthanded, $\omega > 0 \Rightarrow \tau^-$ righthanded).
The data (points with errors) as well as the Monte Carlo 
expectation for $h_{\nu_\tau}\xi =-1$ (solid histograms) 
show clearly the spin correlation, 
whereas for $h_{\nu_\tau}\xi=0$ (dashed histograms) 
the two sides of the event are uncorrelated. 
The hatched histograms show
the Monte Carlo predicted background 
(assuming $h_{\nu_\tau}\xi = -1$).
}
\end{center}
\end{figure}

\begin{figure}[thb]
\begin{center}
\leavevmode
\unitlength1.0cm
\epsfxsize=16.5cm
\epsfysize=6.0cm
\epsffile{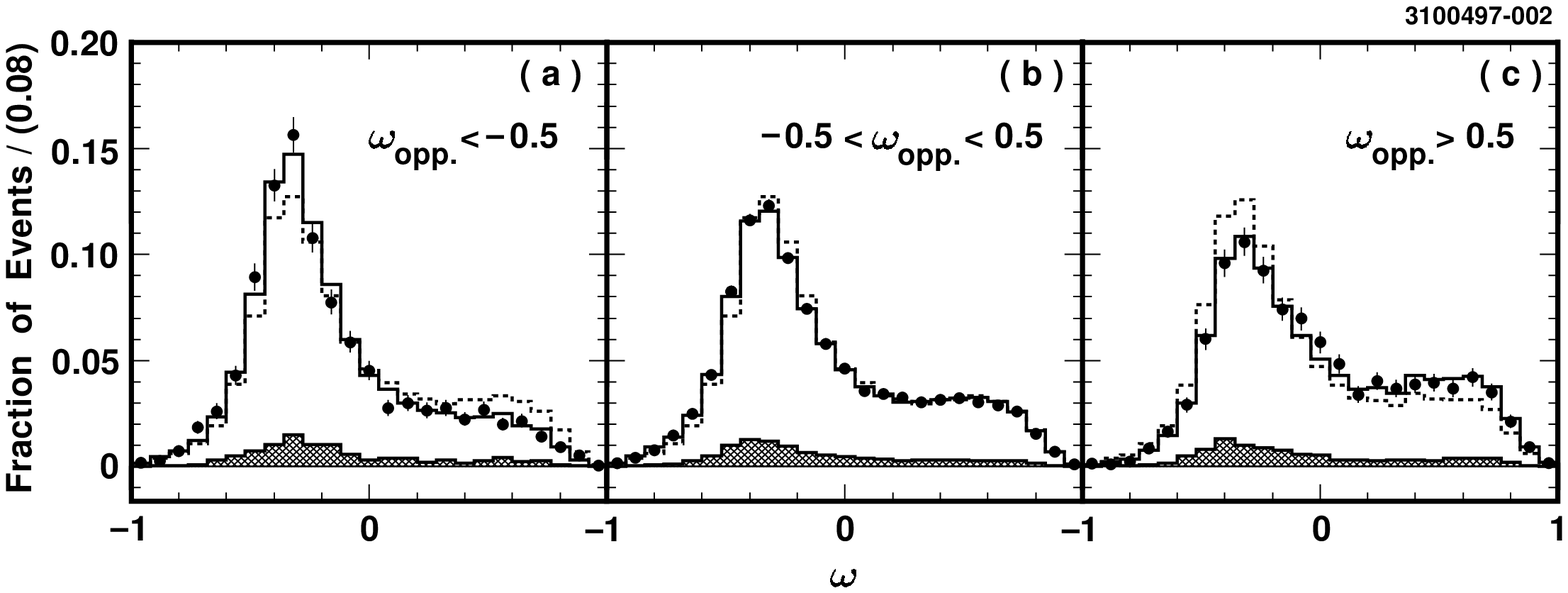}
\caption{\label{sens_rho}
The $\omega$ spectrum for the $\rho$-vs-$\rho$ sample  
of one side of the event for different values of $\omega$
from the other side (two entries per event).
The data are represented by the dots with error bars. The solid
histograms show the Monte Carlo expectation for
$h_{\nu_\tau}^2 =1$. The Monte Carlo
expectation for $h_{\nu_\tau}^2 =0$ is given by the
dashed histograms. Background is indicated by the 
hatched histograms (assuming $h_{\nu_\tau}^2 =1$).
}
\end{center}
\end{figure}

The spin correlation used in this analysis can be most
easily illustrated with the spin sensitive variable
$\omega$ \cite{davier} of the decay
$\tau^\mp\rightarrow\pi^\mp\pi^0\nu$.
Assuming the Standard Model $V$-$A$ interaction,
the matrix element of this decay
in the $\tau$ rest frame is
\begin{equation}
\label{semi}
{\vert {\cal{M}} 
(\tau^\mp\rightarrow\pi^\mp \pi^0 \nu)\vert}^2 =
{\vert \overline{\cal{M}} \vert}^2 ( 1 \pm s_z H_z ) \, ,
\end{equation}
where the spin quantization axis $\hat{z}$ is chosen along
the flight direction of the $\tau$ in the
laboratory frame. The factor 
${\vert \overline{\cal{M}} \vert}^2$ is the spin
averaged matrix element. The spin dependent part of the
matrix element depends on
$H_z$ and $s_z$, where $H_z$ and $s_z$ 
are the $z$-components of the polarimetric vector 
and the $\tau$ spin vector. For the decay 
$\tau^\mp\rightarrow\pi^\mp\pi^0\nu$, $H_\mu$ is given by
\begin{equation}
\label{pol}
H_\mu = 
\frac{ m_\tau ( 2 Q_\mu (kQ) - k_\mu (QQ))}{ 2(kQ)(qQ) - (kq)(QQ)} \, ,
\end{equation}
where $k$ denotes the $\tau$ neutrino momentum, 
$q$ the $\tau$ momentum, and
$Q=p_{\pi^\mp} - p_{\pi^0}$.

As can be seen from Eqn.~\ref{semi}, lefthanded $\tau^-$ leptons
(righthanded $\tau^+$ leptons) have preferentially negative
$H_z$ values, whereas righthanded $\tau^-$ leptons 
(lefthanded $\tau^+$ leptons) have preferentially positive
$H_z$ values.
Averaging $H_z$ over the kinematically allowed $\tau$ rest 
frames yields the variable $\omega$.

Fig.~\ref{sens_elk} shows the dependence of the measured 
electron momentum spectrum on $\omega$. 
One clearly sees that for negative $\omega$
values, high momentum leptons are preferred by the data, 
whereas for positive $\omega$ values, low momentum leptons
are preferred. This correlation indicates 
$h_{\nu_\tau}\xi \approx -1$ as expected by the Standard Model.
For $h_{\nu_\tau}\xi =0$, which is equivalent to zero spin correlation,
the lepton momentum spectrum is independent of $\omega$,
as can be seen in Fig.~\ref{sens_elk}.
The corresponding plots to Fig.~\ref{sens_elk}
for the $\mu$-vs-$\rho$ sample are very similar and are not shown here.
Fig.~\ref{sens_rho} shows for the $\rho$-vs-$\rho$ sample
the $\omega$ spectrum of one side of the event for different values
of $\omega$ from the other side (two entries per event). 
Again the data favor $h_{\nu_\tau}^2 \approx  1$ in agreement
with the Standard Model.

To take the background from other $\tau$ events into 
account, the fit function of Eqn.~\ref{likel} is 
extended to include background
\begin{equation}
L_i = 
\left(  
1 - \left( \alpha_1 +  \ldots + \alpha_n \right)  
\right)S_i 
  + \alpha_1 B_{1,i} + \ldots + \alpha_n B_{n,i} \, ,
\end{equation}
where $\alpha_k$ is the background fraction of the
$k$-th background. The function
$S_i$ is the likelihood of the signal events,
given by Eqn.~\ref{likel}. The functions 
$B_{k,i}$ are the corresponding likelihoods of the 
backgrounds. 
For the dominant sources of background, 
listed in Table \ref{back}, the functions $B_{k,i}$
include their full dependence on the
fit parameters $\Theta$ to avoid bias.
The values used in the fits for the background 
fractions $\alpha_k$ are taken from Table \ref{back}.
The amount of background not included in the
fit is $\approx 1\%$ in the $l$-vs-$\rho$ sample and 
around $2\%$ in the $\rho$-vs-$\rho$ sample. 
The effect of this disregarded background 
(``remaining sources'' in Table \ref{syst})
is discussed in the systematic error section.

The hadronic current of the spin 
analyzer $\tau^\mp\rightarrow\pi^\mp\pi^0\nu$ 
is very well known. However, the $q^2$-dependence of the
intermediate resonance structure might be a possible
source for uncertainties, especially the contribution of the
$\rho^\prime$ meson. The $\rho^\prime$ contribution is
parametrized by $\beta$ \cite{kuhnsant}, and 
with $e^+ e^-$ data \cite{barkov} $\beta$ is determined 
\cite{kuhnsant} to be $\beta =-0.145$.
In a recent CLEO measurement \cite{jon}, 
where $\tau$ events were used, 
a value of $\beta = -0.091\pm 0.009$ is measured.
For consistency, we use the latter value together with the
mass and width obtained in ref.~\cite{jon} for the
$\rho$ and $\rho^\prime$ mesons.

The only fit parameter in the $\rho$-vs-$\rho$ analysis is
$h^2_{\nu_\tau}$. We obtain
\begin{displaymath}
h^2_{\nu_\tau} = 0.989 \pm 0.019 \, .
\end{displaymath}

In the $e$-vs-$\rho$ analysis we have three fit
parameters $\rho$, $h_{\nu_\tau}\xi$, and 
$h_{\nu_\tau}\xi\delta$. 
We measure the following values:
\begin{displaymath}
\rho_e = 0.747\pm 0.012 \quad h_{\nu_\tau}\xi_e = 0.973 \pm 0.047 \quad 
h_{\nu_\tau}\xi_e \delta_e = 0.716 \pm 0.031 
\end{displaymath}
As mentioned in the introduction, the 
$l$-vs-$\rho$ sample used here is correlated with the one
used in ref.~\cite{mandeepa}.
Because of the special treatment of the low energy muons there,
the precision on the Michel parameter $\eta$ reached in 
ref.~\cite{mandeepa} is better than the one we might obtain here.
Therefore, we do not fit for $\eta$.
Instead we fixed
$\eta$ to the value determined in ref.~\cite{mandeepa} of
$\eta_\mu = 0.010 \pm 0.261$ and $\eta_\mu = -0.015\pm 0.091$, where the
first result is obtained in the muon sample alone and 
the second one is the combined result of the muon and electron
sample under the assumption $\rho_e = \rho_\mu$.
With the first result  we obtain in the 
$\mu$-vs-$\rho$ analysis
\begin{displaymath}
\rho_\mu = 0.750\pm 0.017 \quad h_{\nu_\tau}\xi_\mu = 1.048 \pm 0.068 \quad 
h_{\nu_\tau}\xi_\mu\delta_\mu = 0.781 \pm 0.040 \, .
\end{displaymath}
Using the second result
we measure with the $\mu$-vs-$\rho$ sample
\begin{displaymath}
\rho_\mu = 0.746\pm 0.017 \quad h_{\nu_\tau}\xi_\mu  = 1.043 \pm 0.067 \quad 
h_{\nu_\tau}\xi_\mu \delta_\mu = 0.777 \pm 0.040 \, .
\end{displaymath}
All errors shown above are statistical only.

Fixing $\eta$ to the Standard Model value of $\eta=0$ 
results in a shift of 
$\rho_{\eta = -0.015} - \rho_{\eta=0}= -0.0029$, 
$(h_{\nu_\tau}\xi)_{\eta = -0.015} - 
(h_{\nu_\tau}\xi)_{\eta = 0}= -0.0024$, and
$(h_{\nu_\tau}\xi\delta)_{\eta = -0.015} - 
(h_{\nu_\tau}\xi\delta)_{\eta = 0}=-0.0018$.  

\begin{figure}[th]
\begin{center}
\leavevmode
\unitlength1.0cm
\epsfxsize=14.84cm
\epsfysize=15.5cm
\epsffile{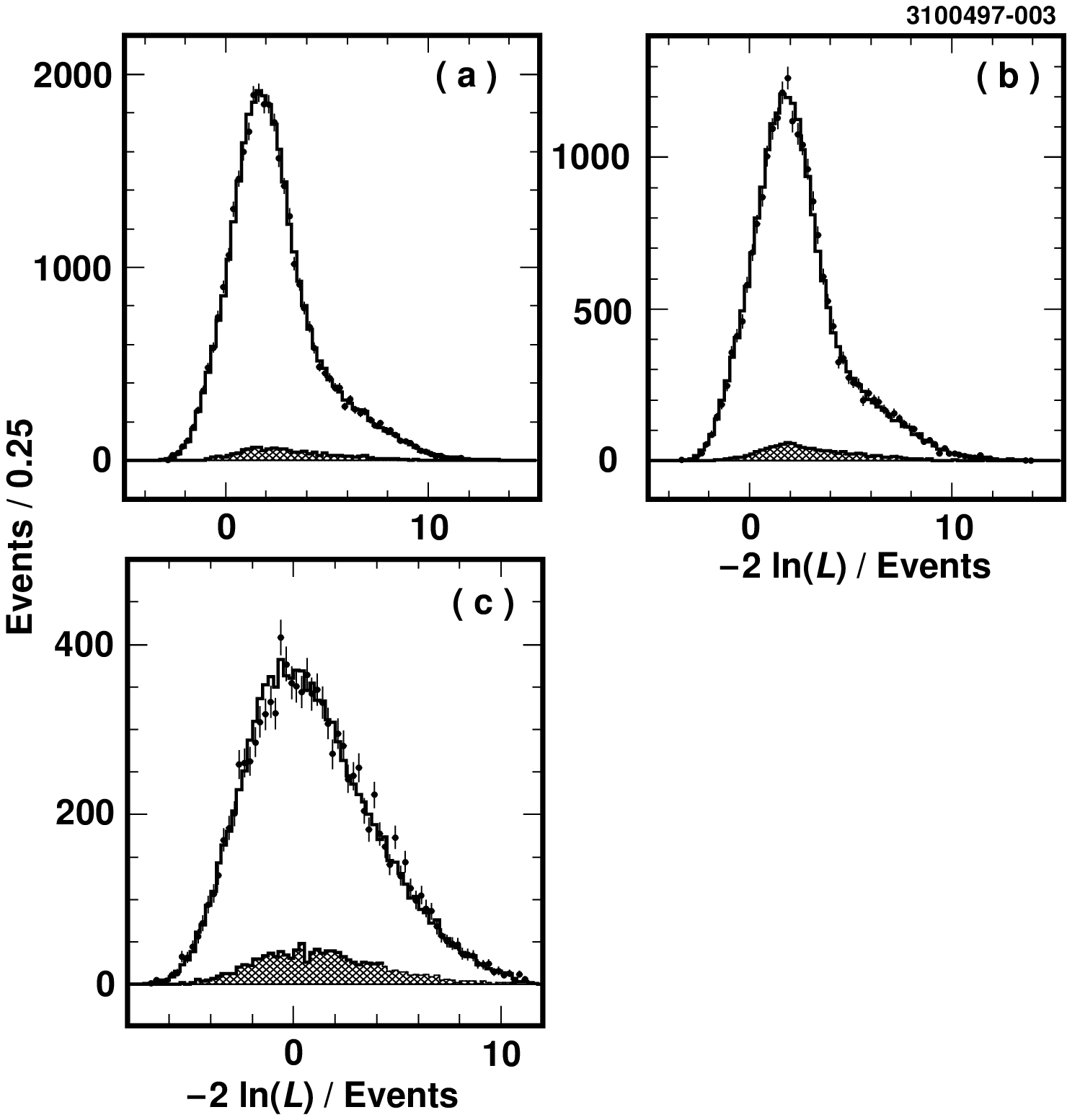}
\caption{\label{guete}
Contributions of single events to 
$-2\ln L$, for (a) 
$(e^\mp \nu\overline{\nu})(\pi^\pm\pi^0 \overline{\nu})$,
(b)  
$(\mu^\mp \nu\overline{\nu})(\pi^\pm\pi^0 \overline{\nu})$, 
and (c)
$(\pi^\mp\pi^0 \nu)(\pi^\pm\pi^0 \overline{\nu})$ events.
Data (dots with error bars) and Monte Carlo 
(solid histograms) are in good agreement. 
Background is represented by the hatched histograms.}
\end{center}
\end{figure}

The confidence levels of  the fits are $73\%$ 
in the $e$-vs-$\rho$ analysis, 
$21\%$ in the $\mu$-vs-$\rho$ analysis, 
and $9\%$ in the $\rho$-vs-$\rho$ analysis. 
Fig.~\ref{guete} shows the corresponding likelihood per 
event distributions. 
One sees that the data are in good agreement 
with the best fit model. 

{\footnotesize
\begin{table}[t]
\caption{Contributions to the systematic error.}
\label{syst}
\begin{center}
\begin{tabular}{l c c c c c c c }
    &$\Delta(\rho_e)$
    &$\Delta(\rho_\mu)$ 
    &$\Delta(h_{\nu_\tau}\xi_e)$
    &$\Delta(h_{\nu_\tau}\xi_\mu)$
    &$\Delta(h_{\nu_\tau}\xi_e \delta_e )$
    &$\Delta(h_{\nu_\tau}\xi_\mu \delta_\mu )$
    &$\Delta(h^2_{\nu_\tau})$  \\ 
\hline
Monte Carlo         & & & & & & & \\ 
\ statistics  & 
\raisebox{1.0ex}[-1.5ex]{$\pm 0.0025$} &
\raisebox{1.0ex}[-1.5ex]{$\pm 0.0028$} &
\raisebox{1.0ex}[-1.5ex]{$\pm 0.0122$} &
\raisebox{1.0ex}[-1.5ex]{$\pm 0.0131$} &
\raisebox{1.0ex}[-1.5ex]{$\pm 0.0090$} &
\raisebox{1.0ex}[-1.5ex]{$\pm 0.0072$} &
\raisebox{1.0ex}[-1.5ex]{$\pm 0.0014$} \\ 
lepton       & & & & & & & \\  
\ identification & 
\raisebox{1.0ex}[-1.5ex]{$\pm 0.0006$} &
\raisebox{1.0ex}[-1.5ex]{$\pm 0.0033$} &
\raisebox{1.0ex}[-1.5ex]{$\pm 0.0005$} &
\raisebox{1.0ex}[-1.5ex]{$\pm 0.0016$} &
\raisebox{1.0ex}[-1.5ex]{$\pm 0.0004$} &
\raisebox{1.0ex}[-1.5ex]{$\pm 0.0025$} &
                                     \\ 
acceptance function  & & & & & & & \\  
\ of spin analyzer & 
\raisebox{1.0ex}[-1.5ex]{$\pm 0.0010$} &
\raisebox{1.0ex}[-1.5ex]{$\pm 0.0015$} &
\raisebox{1.0ex}[-1.5ex]{$\pm 0.0038$} &
\raisebox{1.0ex}[-1.5ex]{$\pm 0.0093$} &
\raisebox{1.0ex}[-1.5ex]{$\pm 0.0036$} &
\raisebox{1.0ex}[-1.5ex]{$\pm 0.0064$} &
\raisebox{1.0ex}[-1.5ex]{$\pm 0.0047$} \\
considered         & & & & & & & \\ 
\ background & 
\raisebox{1.0ex}[-1.5ex]{$\pm 0.0009$} &
\raisebox{1.0ex}[-1.5ex]{$\pm 0.0012$} &
\raisebox{1.0ex}[-1.5ex]{$\pm 0.0018$} &
\raisebox{1.0ex}[-1.5ex]{$\pm 0.0067$} &
\raisebox{1.0ex}[-1.5ex]{$\pm 0.0008$} &
\raisebox{1.0ex}[-1.5ex]{$\pm 0.0023$} &
\raisebox{1.0ex}[-1.5ex]{$\pm 0.0022$} \\ 
disregarded         & & & & & & & \\ 
\ background &
\raisebox{1.0ex}[-1.5ex]{$\pm 0.0004$} &
\raisebox{1.0ex}[-1.5ex]{$\pm 0.0009$} &
\raisebox{1.0ex}[-1.5ex]{$\pm 0.0029$} &
\raisebox{1.0ex}[-1.5ex]{$\pm 0.0059$} &
\raisebox{1.0ex}[-1.5ex]{$\pm 0.0011$} &
\raisebox{1.0ex}[-1.5ex]{$\pm 0.0017$} &
\raisebox{1.0ex}[-1.5ex]{$\pm 0.0005$} \\ 
parameter $\beta$         & & & & & & & \\ 
\ of $\rho^\prime$ contribution & 
\raisebox{1.0ex}[-1.5ex]{$\pm 0.0002$} &
\raisebox{1.0ex}[-1.5ex]{$\pm 0.0002$} &
\raisebox{1.0ex}[-1.5ex]{$\pm 0.0012$} &
\raisebox{1.0ex}[-1.5ex]{$\pm 0.0025$} &
\raisebox{1.0ex}[-1.5ex]{$\pm 0.0002$} &
\raisebox{1.0ex}[-1.5ex]{$\pm 0.0002$} &
\raisebox{1.0ex}[-1.5ex]{$\pm 0.0010$} \\ 
Michel parameter $\eta$    & & & & & & & \\ 
\  $\Delta(\eta_\mu ) =     \pm 0.091$ \cite{mandeepa} & 
                                      &
\raisebox{1.0ex}[-1.5ex]{$\pm 0.0170$} &
                                      &
\raisebox{1.0ex}[-1.5ex]{$\pm 0.0144$} &
                                      &
\raisebox{1.0ex}[-1.5ex]{$\pm 0.0125$} &
                                       \\ 
Michel parameter $\eta$        & & & & & & & \\ 
\  $\Delta(\eta_\mu ) = \pm 0.261$ \cite{mandeepa} & 
                                      &
\raisebox{1.0ex}[-1.5ex]{$\pm 0.0448$} &
                                      &
\raisebox{1.0ex}[-1.5ex]{$\pm 0.0417$} &
                                      &
\raisebox{1.0ex}[-1.5ex]{$\pm 0.0302$} &
                                       \\ 
detector         & & & & & & & \\ 
\ resolution & 
\raisebox{1.0ex}[-1.5ex]{$\pm 0.0004$} &
\raisebox{1.0ex}[-1.5ex]{$\pm 0.0004$} &
\raisebox{1.0ex}[-1.5ex]{$\pm 0.0006$} &
\raisebox{1.0ex}[-1.5ex]{$\pm 0.0005$} &
\raisebox{1.0ex}[-1.5ex]{$\pm 0.0003$} &
\raisebox{1.0ex}[-1.5ex]{$\pm 0.0004$} &
\raisebox{1.0ex}[-1.5ex]{$\pm 0.0002$} \\
radiation    & 
$\pm 0.0013$ &
$\pm 0.0011$ &
$\pm 0.0018$ &
$\pm 0.0032$ &
$\pm 0.0021$ &
$\pm 0.0041$ &
$\pm 0.0007$ \\ 
trigger      & 
$\pm 0.0017$ &
$\pm 0.0035$ &
$\pm 0.0094$ &
$\pm 0.0123$ &
$\pm 0.0022$ &
$\pm 0.0025$ &
$\pm 0.0011$ \\ 
\hline
total          & & & & & & & \\ 
\  $\Delta(\eta_\mu ) =     \pm 0.091$ \cite{mandeepa} & 
\raisebox{1.0ex}[-1.5ex]{$\pm 0.004$} & 
\raisebox{1.0ex}[-1.5ex]{$\pm 0.018$} &
\raisebox{1.0ex}[-1.5ex]{$\pm 0.016$} & 
\raisebox{1.0ex}[-1.5ex]{$\pm 0.027$} &
\raisebox{1.0ex}[-1.5ex]{$\pm 0.010$} & 
\raisebox{1.0ex}[-1.5ex]{$\pm 0.017$} &
\raisebox{1.0ex}[-1.5ex]{$\pm 0.006$} \\ 
total          & & & & & & & \\ 
\  $\Delta(\eta_\mu ) = \pm 0.261$ \cite{mandeepa} & 
\raisebox{1.0ex}[-1.5ex]{$\pm 0.004$} &
\raisebox{1.0ex}[-1.5ex]{$\pm 0.045$} &
\raisebox{1.0ex}[-1.5ex]{$\pm 0.016$} &
\raisebox{1.0ex}[-1.5ex]{$\pm 0.047$} &
\raisebox{1.0ex}[-1.5ex]{$\pm 0.010$} &
\raisebox{1.0ex}[-1.5ex]{$\pm 0.032$} &
\raisebox{1.0ex}[-1.5ex]{$\pm 0.006$} \\ 
\end{tabular} 
\end{center}
\end{table}
}

Systematic errors arise from statistical errors of the
Monte Carlo estimate of the normalization integral,
momentum dependence of the lepton identification efficiency,
the acceptance function of the $\pi^\mp\pi^0$ spin analyzer,
background, the model for the hadronic current, 
detector resolution, radiation, and trigger.
The different contributions to the systematic error
are summarized in Table \ref{syst}.

The lepton identification efficiency has been measured, as
a function of momentum and polar angle, with independent lepton
data samples. The systematic error given in Table \ref{syst}
arises from the statistical error of this measurement.
The acceptance function of the $\pi^\mp\pi^0$ spin analyzer
has been varied via its dependence on the 
momenta of the two pions and the angle between the two pions. 
The variations considered have been determined 
by comparison of Monte Carlo and data.
The systematic error due to the considered background has
been evaluated by varying the fractions of the different 
backgrounds in the fit function over a range 
given by statistical, experimental, and theoretical uncertainties.
The effect of the disregarded background has been studied
using Monte Carlo. 
The $\rho^\prime$ contribution is measured in ref.~\cite{jon}
with an error of $\Delta\beta= \pm 0.009$. Since $\beta$ has
model dependencies, and to be conservative,
we varied $\beta$ in the range of $\pm 0.020$.
The systematic error due to $\eta$ has been evaluated by 
varying $\eta$ in its determined range \cite{mandeepa}.
The uncertainty in the detector resolution has been estimated
by scaling the error matrix of the resolution by a factor of four.
The systematic error due to radiation has been obtained 
by varying the amount of radiation in the fit function
by $\pm 10\%$.
The uncertainty in the trigger efficiency arises from the
tracking component of the trigger, whereas the uncertainty
due to the neutral component of the trigger is negligible.
Therefore, the systematic error due to the trigger has been
evaluated with a subsample of our data that satisfies
the neutral as well as the tracking component of the trigger.

\begin{table}[t]
\caption{Results. $\rho$, $\xi$, and $\xi\delta$ denote
the combined results.
$\rho_l$, $\xi_l$, and $\xi_l\delta_l$ are the results
separated for electrons and muons.}
\label{res}
\begin{center}
\begin{tabular}{l r@{}l r@{}l@{}l l@{}c@{}l l@{}c@{}l}
& 
\multicolumn{2}{c}{world average \cite{pdg}} 
&
\multicolumn{3}{c}{this analysis$^*$}
& & & & & &  \\ \cline{1-6}
$ h_{\nu_\tau}                                  $ &
$ -1.011                                        $ & 
$ \pm 0.027                                     $ &
$ -0.995                                        $ & 
$ \pm 0.010                                     $ & 
$ \pm 0.003                                     $ & 
\multicolumn{6}{c}{%
\raisebox{1.4ex}[-1.4ex]{correlation coefficients}} \\ 
$ \rho                                          $ &
$ 0.742                                         $ & 
$ \pm 0.027                                     $ &
$ 0.747                                         $ & 
$ \pm 0.010                                     $ & 
$ \pm 0.006                                     $ & 
\raisebox{1.4ex}[-1.4ex]{$ \kappa(\rho , \xi )  $}& 
\raisebox{1.4ex}[-1.4ex]{$ =                          $}& 
\raisebox{1.4ex}[-1.4ex]{$ \phantom{-} 0.046          $}& 
\raisebox{1.4ex}[-1.4ex]{$ \kappa(\rho , \xi\delta )  $}& 
\raisebox{1.4ex}[-1.4ex]{$ =                    $}& 
\raisebox{1.4ex}[-1.4ex]{$ \phantom{-} 0.069    $}           \\
$ \xi                                           $ &
$ 1.03 \phantom{0}                              $ & 
$ \pm 0.12                                      $ &
$ 1.007                                         $ & 
$ \pm 0.040                                     $ & 
$ \pm 0.015                                     $ &
\raisebox{1.4ex}[-1.4ex]{$ \kappa(\rho ,h_{\nu_\tau}  ) $}& 
\raisebox{1.4ex}[-1.4ex]{$ =                            $}& 
\raisebox{1.4ex}[-1.4ex]{$ \phantom{-} 0.000            $}& 
\raisebox{1.4ex}[-1.4ex]{$ \kappa(\xi , \xi\delta )     $}& 
\raisebox{1.4ex}[-1.4ex]{$ =                    $}& 
\raisebox{1.4ex}[-1.4ex]{$ \phantom{-} 0.158    $}           \\
$ \xi\delta                                     $ & 
$ 0.76 \phantom{0}                              $ & 
$ \pm 0.11                                      $ &    
$ 0.745                                         $ & 
$ \pm 0.026                                     $ & 
$ \pm 0.009                                     $ & 
\raisebox{1.4ex}[-1.4ex]{$ \kappa(\xi ,h_{\nu_\tau}  )        $}& 
\raisebox{1.4ex}[-1.4ex]{$ =                                  $}& 
\raisebox{1.4ex}[-1.4ex]{$ -0.241                             $}& 
\raisebox{1.4ex}[-1.4ex]{$ \kappa(\xi\delta  ,h_{\nu_\tau} ) $}& 
\raisebox{1.4ex}[-1.4ex]{$ =                                  $}& 
\raisebox{1.4ex}[-1.4ex]{$ -0.276                             $}\\
$ \rho_e                                        $ &
$ 0.736                                         $ & 
$ \pm 0.028                                     $ &
$ 0.747                                         $ & 
$ \pm 0.012                                     $ & 
$ \pm 0.004                                     $ & 
\raisebox{1.4ex}[-1.4ex]{$ \kappa(\rho_e , \xi_e )         $}& 
\raisebox{1.4ex}[-1.4ex]{$ =                               $}& 
\raisebox{1.4ex}[-1.4ex]{$ \phantom{-} 0.046               $}& 
\raisebox{1.4ex}[-1.4ex]{$ \kappa(\rho_e , \xi_e\delta_e ) $}& 
\raisebox{1.4ex}[-1.4ex]{$ =                               $}& 
\raisebox{1.4ex}[-1.4ex]{$ \phantom{-} 0.074               $}\\
$ \xi_e                                         $ &
$ 1.03 \phantom{0}                              $ & 
$ \pm 0.25                                      $ &
$ 0.979                                         $ & 
$ \pm 0.048                                     $ & 
$ \pm 0.016                                     $ & 
\raisebox{1.4ex}[-1.4ex]{$ \kappa(\rho_e ,h_{\nu_\tau}  ) $}& 
\raisebox{1.4ex}[-1.4ex]{$ =                              $}& 
\raisebox{1.4ex}[-1.4ex]{$ \phantom{-} 0.000              $}& 
\raisebox{1.4ex}[-1.4ex]{$ \kappa(\xi_e , \xi_e\delta_e ) $}& 
\raisebox{1.4ex}[-1.4ex]{$ =                              $}& 
\raisebox{1.4ex}[-1.4ex]{$ \phantom{-} 0.216              $}\\
$ \xi_e\delta_e                                 $ & 
$ 1.11 \phantom{0}                              $ & 
$ \pm 0.18                                      $ &    
$ 0.720                                         $ & 
$ \pm 0.032                                     $ & 
$ \pm 0.010                                     $ & 
\raisebox{1.4ex}[-1.4ex]{$ \kappa(\xi_e ,h_{\nu_\tau}  )          $}& 
\raisebox{1.4ex}[-1.4ex]{$ =                                      $}& 
\raisebox{1.4ex}[-1.4ex]{$ -0.194                                 $}& 
\raisebox{1.4ex}[-1.4ex]{$ \kappa(\xi_e\delta_e ,h_{\nu_\tau} ) $}& 
\raisebox{1.4ex}[-1.4ex]{$ =                                      $}& 
\raisebox{1.4ex}[-1.4ex]{$ -0.214                                 $}\\
$ \rho_\mu                                      $ &
$ 0.74                                          $ & 
$ \pm 0.04                                      $ &
$ 0.750                                         $ & 
$ \pm 0.017                                     $ & 
$ \pm 0.045                                     $ & 
\raisebox{1.4ex}[-1.4ex]{$ \kappa(\rho_\mu , \xi_\mu )           $}& 
\raisebox{1.4ex}[-1.4ex]{$ =                                     $}& 
\raisebox{1.4ex}[-1.4ex]{$ \phantom{-} 0.026                     $}& 
\raisebox{1.4ex}[-1.4ex]{$ \kappa(\rho_\mu , \xi_\mu\delta_\mu ) $}& 
\raisebox{1.4ex}[-1.4ex]{$ =                                     $}& 
\raisebox{1.4ex}[-1.4ex]{$ \phantom{-} 0.029                     $}\\
$ \xi_\mu                                       $ &
$ 1.23 \phantom{0}                              $ & 
$ \pm 0.24                                      $ &
$ 1.054                                         $ & 
$ \pm 0.069                                     $ & 
$ \pm 0.047                                     $ & 
\raisebox{1.4ex}[-1.4ex]{$ \kappa(\rho_\mu ,h_{\nu_\tau}  )      $}& 
\raisebox{1.4ex}[-1.4ex]{$ =                                     $}& 
\raisebox{1.4ex}[-1.4ex]{$ \phantom{-} 0.000                     $}& 
\raisebox{1.4ex}[-1.4ex]{$ \kappa(\xi_\mu , \xi_\mu\delta_\mu )  $}& 
\raisebox{1.4ex}[-1.4ex]{$ =                                     $}& 
\raisebox{1.4ex}[-1.4ex]{$ - 0.030                               $}\\
$ \xi_\mu\delta_\mu                             $ & 
$ 0.71 \phantom{0}                              $ & 
$ \pm 0.15                                      $ &    
$ 0.786                                         $ & 
$ \pm 0.041                                     $ & 
$ \pm 0.032                                     $ & 
\raisebox{1.4ex}[-1.4ex]{$ \kappa(\xi_\mu ,h_{\nu_\tau}  )            $}& 
\raisebox{1.4ex}[-1.4ex]{$ =                                          $}& 
\raisebox{1.4ex}[-1.4ex]{$ -0.128                                     $}& 
\raisebox{1.4ex}[-1.4ex]{$ \kappa(\xi_\mu\delta_\mu ,h_{\nu_\tau} ) $}& 
\raisebox{1.4ex}[-1.4ex]{$ =                                          $}& 
\raisebox{1.4ex}[-1.4ex]{$ -0.152                                     $}\\
\multicolumn{12}{l}{%
\footnotesize $^*$together with the sign of $h_{\nu_\tau}$ 
determined in \cite{albin2,ich}} \\
\end{tabular} 
\end{center}
\end{table}

With these systematic errors added in quadrature and 
using the sign of $h_{\nu_\tau}$ determined 
in ref.~\cite{albin2,ich} we obtain the results 
listed in Table \ref{res}. The results are in agreement 
with the Standard $V$-$A$ interaction.


\section{INTERPRETATION AND SUMMARY}

\begin{figure}[thb]
\begin{center}
\leavevmode
\unitlength1.0cm
\epsfxsize=13.5cm
\epsfysize=18.0cm
\epsffile{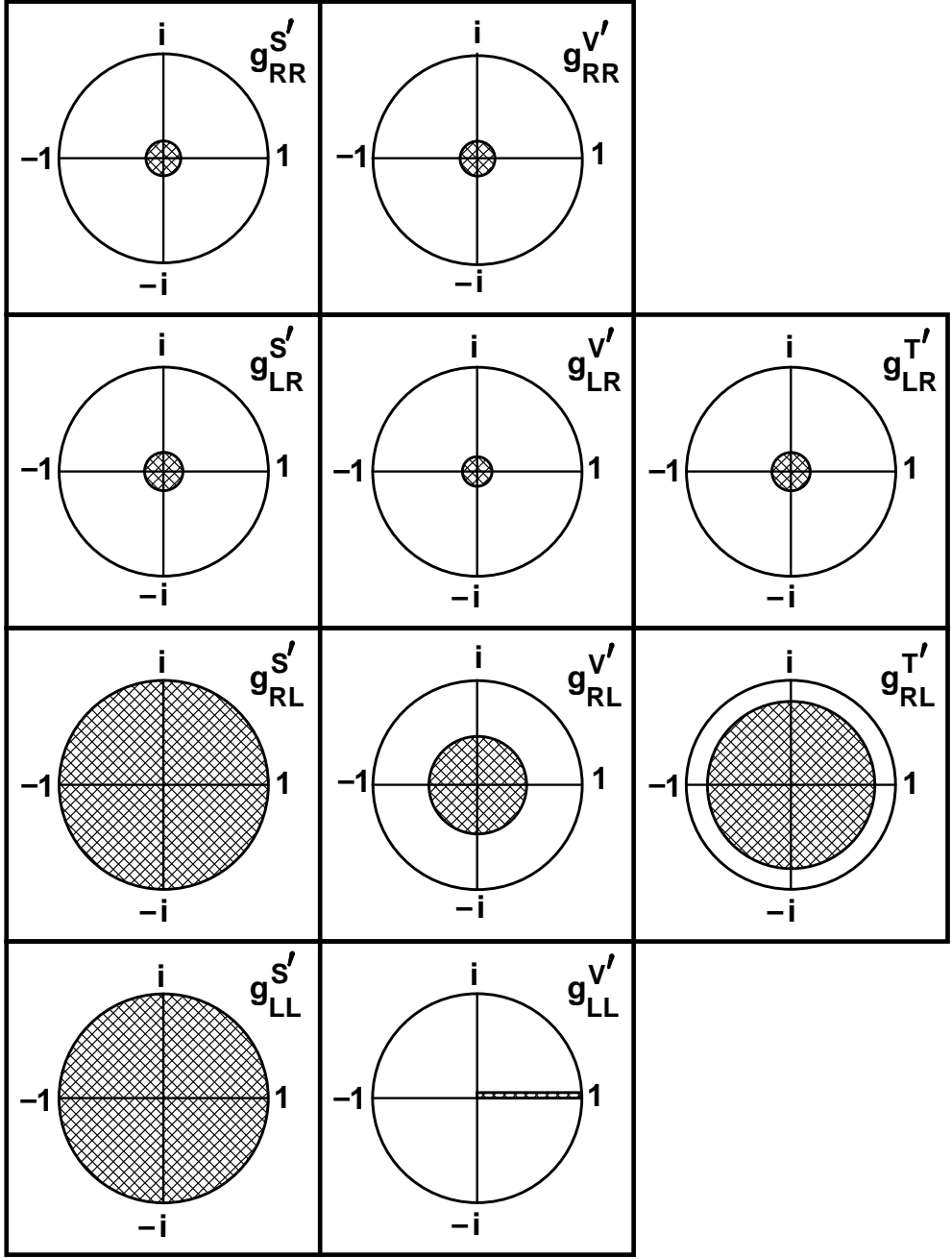}
\caption{\label{coupl}
$90\%$ confidence limits on the reduced coupling constants 
$g^{\gamma^\prime}_{\epsilon\mu}= 
g^\gamma_{\epsilon\mu}/
max(g^\gamma_{\epsilon\mu})$. 
}
\end{center}
\end{figure}

\begin{figure}[thb]
\begin{center}
\leavevmode
\unitlength1.0cm
\epsfxsize=14.cm
\epsfysize=8.0cm
\epsffile{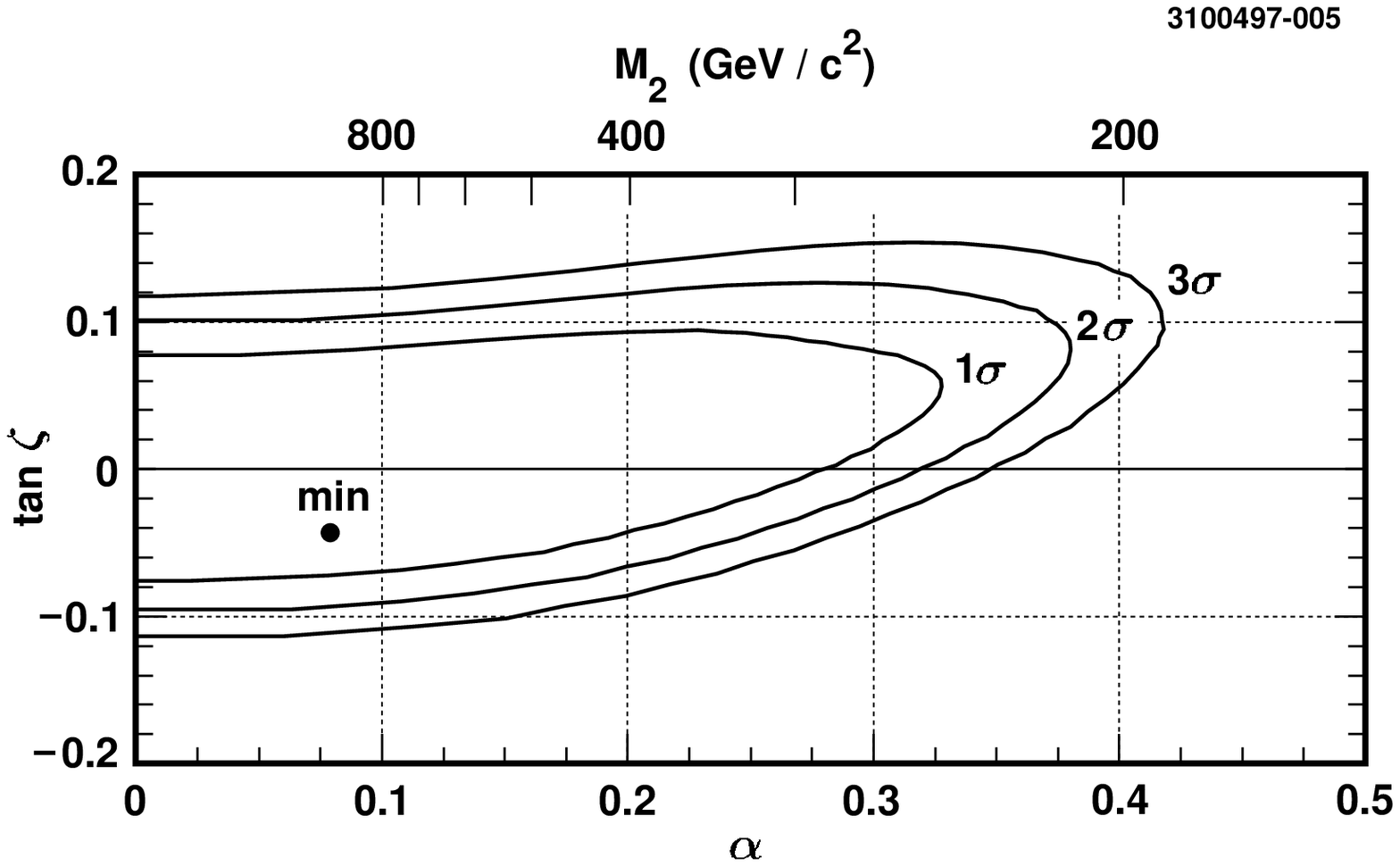}
\caption{\label{cont}}
Limits on the mass ratio $\alpha$ and the mixing angle
$\zeta$ of a left-right symmetric model.
\end{center}
\end{figure}

\begin{figure}[thb]
\begin{center}
\leavevmode
\unitlength1.0cm
\epsfxsize=14.cm
\epsfysize=8.0cm
\epsffile{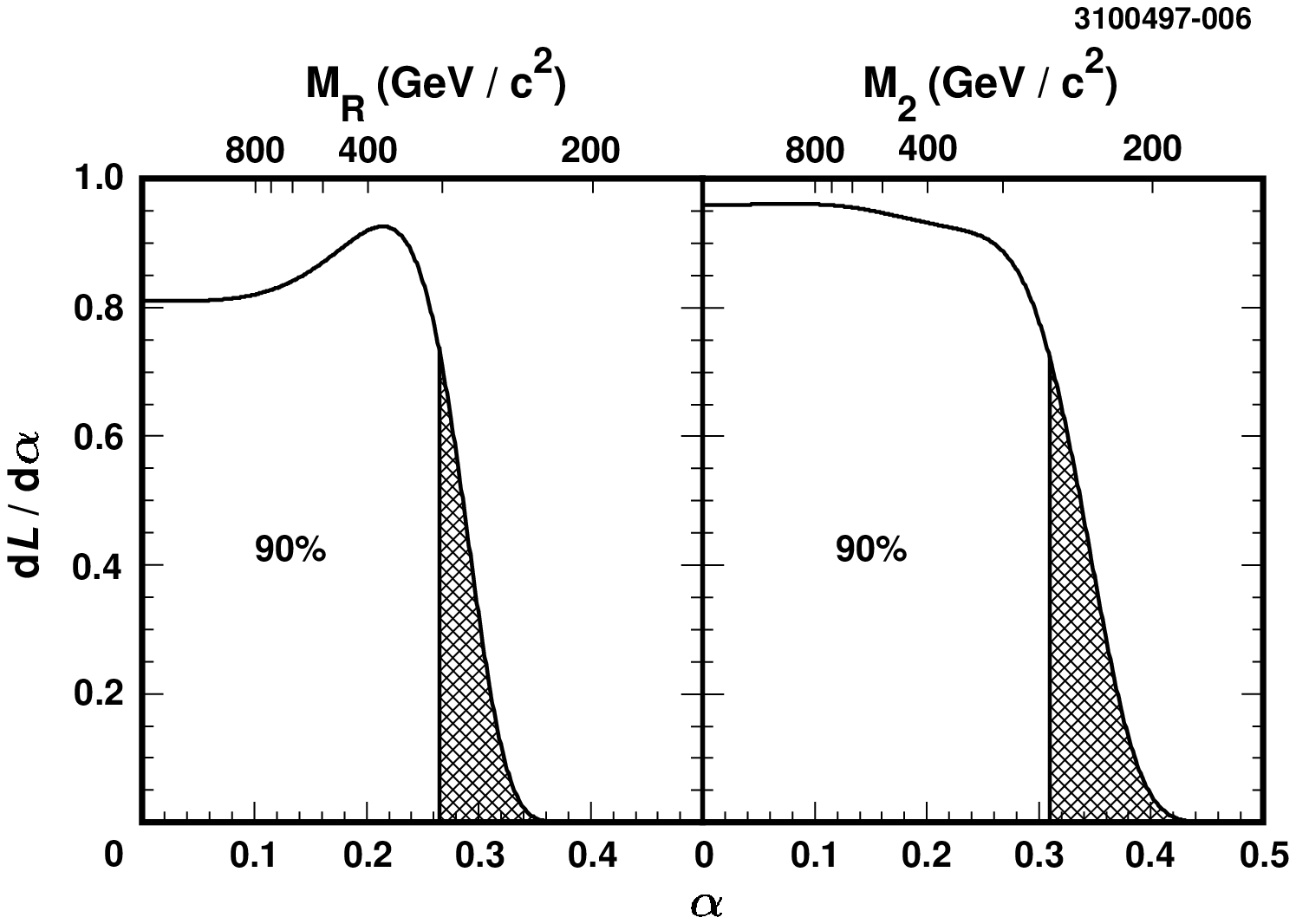}
\caption{\label{lim}}
$90\%$ confidence limits on the mass ratio $\alpha$ for (a)
$\tan\zeta = 0$  and (b) for $\tan\zeta$ free.
\end{center}
\end{figure}

The measurement of $\xi$ and $\delta$ implies that
the probability $P^\tau_R$ (see Eqn.~\ref{rechts})
of a righthanded $\tau$ to participate in
leptonic $\tau$ decays is $P_\tau^R < 0.044$ at a $90\%$ 
confidence level.
Separately for electrons and muons, we obtain 
$P_\tau^R < 0.066$ for the electronic mode and
$P_\tau^R < 0.067$ for the muonic mode.
Both limits are at a $90\%$ confidence level.

The $90\%$ confidence limits on the 
reduced coupling constants 
$g^{\gamma^\prime}_{\epsilon\mu}= 
g^\gamma_{\epsilon\mu}/
max(g^\gamma_{\epsilon\mu})$
obtained from the combined results on the Michel parameters
(Table \ref{res}) are plotted in Fig.~\ref{coupl}.
Without measuring the helicity of the $\tau$ 
neutrino in leptonic decays, the $g^S_{LL}$
coupling cannot be distinguished from the $g^V_{LL}$ 
coupling. Adding the knowledge of the 
parameter $\eta$ does not improve the limits.
The couplings with a righthanded $\tau$, 
$g^\gamma_{\epsilon R}$, are mostly constrained by the
determination of $\xi$ and $\delta$. Additional
information comes from the measurement of $\rho$,
which allows one to constrain the $g^V_{RL}$ and
$g^T_{RL}$ couplings.
Compared to the situation five years ago when
only the Michel parameter $\rho$ was measured
in $\tau$ decay and no limits on the coupling
constants $g^\gamma_{\epsilon\mu}$ existed,
Fig.~\ref{coupl} illustrates the progress made.
However, the $V$-$A$ interaction as assumed by
the Standard Model is still not fully experimentally
verified for $\tau$ decays.

More stringent limits can be obtained by restricting
the generality of the model. For example, we consider a
left-right symmetric model \cite{beg} for the
electroweak interaction, where the parity
violation has its origin in a spontanous
symmetry breaking of the left-right symmetry.
In addition to the pure lefthanded $W$ boson
$W_L$ of the Standard Model, such a model
assumes a pure righthanded $W$ boson $W_R$,
where the mass eigenstates $W_1$ and $W_2$
are in general superpositions of the weak eigenstates
$W_L$ and $W_R$. 
This model can be parametrized by the
mass ratio $\alpha = M_1/M_2$ of the two
bosons $W_{1/2}$ and the mixing angle $\zeta$
between $W_{L/R}$. The Standard Model is
obtained in the limit $\alpha \rightarrow 0$ and
$\zeta\rightarrow 0$.

Fig.~\ref{cont} shows the one, two, and three $\sigma$
contours for $\alpha$ and $\zeta$ obtained with
the combined results on $\rho$, $\xi$, $\xi\delta$,
and $h_{\nu_\tau}$.
For $\zeta = 0$, $W_2$ is identical with $W_R$ and
the following limit is obtained
on $M_R$:
\begin{displaymath} 
M_R > 304 \mbox{ GeV/c}^2  \mbox{ at 90\% CL}
\end{displaymath}
The mass limit obtained for $\zeta$ free is:
\begin{displaymath}
M_2 > 260 \mbox{ GeV/c}^2  \mbox{ at 90\% CL}
\end{displaymath}
The corresponding likelihood functions are shown in Fig.~\ref{lim}.
The limit obtained in  muon decay is 
$M_2 > 406 \mbox{ GeV/c}^2$ \cite{pdg}.

We have also studied the constraints given 
by our measurement on extensions of the Standard Model 
with charged Higgs bosons.
The $\tau^-$ lepton and the charged daughter lepton $l^-$
in leptonic $\tau$ decays mediated by charged Higgs bosons
are righthanded \cite{haber1,haber2}. Thus, in the general ansatz
of Eqn.~\ref{general}, charged Higgs bosons are represented
by the $g^S_{RR}$ coupling. From our measurement on 
$\xi_\mu$ and $\xi_\mu\delta_\mu$ we obtain
\begin{displaymath}
M_{H^\pm} > 0.91\times\tan\beta \mbox{ GeV/c}^2  \mbox{ at 90\% CL ,}
\end{displaymath}
where $\beta$ is the ratio of vacuum expectation values.
The combined limit obtained from this measurement and
the recent CLEO measurement of $\eta$ \cite{mandeepa} is:
\begin{displaymath}
M_{H^\pm} > 1.04\times\tan\beta \mbox{ GeV/c}^2  \mbox{ at 90\% CL}
\end{displaymath}

All results presented here assume massless $\tau$
neutrinos.  
We have sudied the kinematical and dynamical
effects of a $24 \mbox{ MeV/c}^2$ $\tau$ neutrino on
our results and found that for the Michel parameters
as well as for the $\tau$ neutrino helicity
such a neutrino does not affect the results
at the level of our accuracy.

We have presented a precision measurement of the Michel 
parameters $\rho$, $\xi$, and $\delta$ as well as of the
$\tau$ neutrino helicity $h_{\nu_\tau}$.
The results obtained are consistent with
the Standard Model prediction. 
With the exception of the Michel parameter $\eta$,
the CLEO measurement given in ref.~\cite{mandeepa} 
is superseded by the results obtained here.
Despite the high statistics used, the accuracy of
the measurements is still dominated by statistical
and not systematic uncertainties, leaving a
potential of improving the accuracy of the determination of the
Michel parameters in $\tau$ decays at the
$B$ factories soon to come into operations,
as well as at future $\tau$ factories.


\section*{ACKNOWLEDGEMENTS}


We gratefully acknowledge the effort of the CESR staff in 
providing us with excellent luminosity 
and running conditions.
J.P.A., J.R.P., and I.P.J.S. thank                                           
the NYI program of the NSF, 
M.S. thanks the PFF program of the NSF,
G.E. thanks the Heisenberg Foundation, 
%
%
K.K.G., M.S., H.N.N., T.S., and H.Y. thank the
OJI program of DOE, 
J.R.P., K.H., M.S. and V.S. thank the A.P. Sloan Foundation,
A.W. and R.W. thank the 
Alexander von Humboldt Stiftung,
and M.S. thanks Research Corporation
for support.
This work was supported by the National Science Foundation, the
U.S. Department of Energy, and the Natural Sciences 
and Engineering Research Council of Canada.


\end{document}